\documentstyle[10pt,epsf,epsfig,dp_delphititle]{dp_delphi}
\makeindex
\def\DpPaperGroup{PH-EP}
\def\DpPaperRef{2007-010}
\def\DpDate{16 April 2007}
\def\DpAuthors{DELPHI Collaboration}
\def\DpSubmit{(Accepted by Phys. Lett. B)}
\def\DpTitle{{
Measurement of the Tau Lepton Polarisation at LEP2
}}
\def\DpComment{}
\def\DpEMail{}

\begin{document}
%%%%%%%%%%%%%%%%%%%%%%%%%% Is there a problem with Coll.Sty ?
\makeatletter
%\input{dp_system:coll.sty}
% Collapse citation numbers to ranges.  Non-numeric and undefined labels
% are handled.  No sorting is done.  E.g., 1,3,2,3,4,5,foo,1,2,3,?,4,5
% gives 1,3,2-5,foo,1-3,?,4,5
\newcount\@tempcntc
\def\@citex[#1]#2{\if@filesw\immediate\write\@auxout{\string\citation{#2}}\fi
  \@tempcnta\z@\@tempcntb\m@ne\def\@citea{}\@cite{\@for\@citeb:=#2\do
    {\@ifundefined
       {b@\@citeb}{\@citeo\@tempcntb\m@ne\@citea\def\@citea{,}{\bf ?}\@warning
       {Citation `\@citeb' on page \thepage \space undefined}}%
    {\setbox\z@\hbox{\global\@tempcntc0\csname b@\@citeb\endcsname\relax}%
     \ifnum\@tempcntc=\z@ \@citeo\@tempcntb\m@ne
       \@citea\def\@citea{,}\hbox{\csname b@\@citeb\endcsname}%
     \else
      \advance\@tempcntb\@ne
      \ifnum\@tempcntb=\@tempcntc
      \else\advance\@tempcntb\m@ne\@citeo
      \@tempcnta\@tempcntc\@tempcntb\@tempcntc\fi\fi}}\@citeo}{#1}}
\def\@citeo{\ifnum\@tempcnta>\@tempcntb\else\@citea\def\@citea{,}%
  \ifnum\@tempcnta=\@tempcntb\the\@tempcnta\else
   {\advance\@tempcnta\@ne\ifnum\@tempcnta=\@tempcntb \else \def\@citea{--}\fi
    \advance\@tempcnta\m@ne\the\@tempcnta\@citea\the\@tempcntb}\fi\fi}
 
\makeatother
%%%%%%%%%%%%%%%%%%%%%%%%%% ??????????????????????????????????
% Generate the title page
\begin{titlepage}
\pagenumbering{roman}
\CERNpreprint{\DpPaperGroup}{\DpPaperRef} % Reference of the paper
\date{{\small\DpDate}} % Date of the paper
\title{\DpTitle} % Title of the paper
\address{\DpAuthors} % General name of the author(s)

\begin{shortabs} % Start the abstract
\noindent
A first measurement of the average polarisation $P_\tau$
of tau leptons produced in e$^+$e$^-$ annihilation
at energies significantly above the Z resonance
is presented.
The polarisation is determined from the kinematic
spectra of tau hadronic decays.
The measured value $P_\tau = -0.164 \pm 0.125$ 
is consistent with the Standard Model prediction
for the mean LEP energy of 197 GeV.

\end{shortabs}

\vfill

\begin{center}
\DpSubmit \ \\ % Horrible hack to allow to have DpSubmit empty
\DpComment \ \\
\DpEMail \ \\
\end{center}
\vfill
\clearpage

\headsep 10.0pt

\addtolength{\textheight}{10mm}
\addtolength{\footskip}{-5mm}
\begingroup
% Commands to process the author names
%
\newcommand{\DpName}[2]{\hbox{#1$^{\ref{#2}}$},\hfill}
\newcommand{\DpNameTwo}[3]{\hbox{#1$^{\ref{#2},\ref{#3}}$},\hfill}
\newcommand{\DpNameThree}[4]{\hbox{#1$^{\ref{#2},\ref{#3},\ref{#4}}$},\hfill}
\newskip\Bigfill \Bigfill = 0pt plus 1000fill
\newcommand{\DpNameLast}[2]{\hbox{#1$^{\ref{#2}}$}\hspace{\Bigfill}}
%
%\small
\footnotesize
\noindent
\DpName{J.Abdallah}{LPNHE}
\DpName{P.Abreu}{LIP}
\DpName{W.Adam}{VIENNA}
\DpName{P.Adzic}{DEMOKRITOS}
\DpName{T.Albrecht}{KARLSRUHE}
\DpName{R.Alemany-Fernandez}{CERN}
\DpName{T.Allmendinger}{KARLSRUHE}
\DpName{P.P.Allport}{LIVERPOOL}
\DpName{U.Amaldi}{MILANO2}
\DpName{N.Amapane}{TORINO}
% ...
\normalsize
\endgroup
\newpage

\titlefoot{Department of Physics and Astronomy, Iowa State
     University, Ames IA 50011-3160, USA
    \label{AMES}}
\titlefoot{IIHE, ULB-VUB,
     Pleinlaan 2, B-1050 Brussels, Belgium
    \label{BRUSSELS}}
\titlefoot{Physics Laboratory, University of Athens, Solonos Str.
     104, GR-10680 Athens, Greece
    \label{ATHENS}}
\titlefoot{Department of Physics, University of Bergen,
     All\'egaten 55, NO-5007 Bergen, Norway
    \label{BERGEN}}
\titlefoot{Dipartimento di Fisica, Universit\`a di Bologna and INFN,
     Via Irnerio 46, IT-40126 Bologna, Italy
    \label{BOLOGNA}}
\titlefoot{Centro Brasileiro de Pesquisas F\'{\i}sicas, rua Xavier Sigaud 150,
     BR-22290 Rio de Janeiro, Brazil
    \label{BRASIL-CBPF}}
\titlefoot{Inst. de F\'{\i}sica, Univ. Estadual do Rio de Janeiro,
     rua S\~{a}o Francisco Xavier 524, Rio de Janeiro, Brazil
    \label{BRASIL-IFUERJ}}
\titlefoot{Coll\`ege de France, Lab. de Physique Corpusculaire, IN2P3-CNRS,
     FR-75231 Paris Cedex 05, France
    \label{CDF}}
\titlefoot{CERN, CH-1211 Geneva 23, Switzerland
    \label{CERN}}
\titlefoot{Institut de Recherches Subatomiques, IN2P3 - CNRS/ULP - BP20,
     FR-67037 Strasbourg Cedex, France
    \label{CRN}}
\titlefoot{Now at DESY-Zeuthen, Platanenallee 6, D-15735 Zeuthen, Germany
    \label{DESY}}
\titlefoot{Institute of Nuclear Physics, N.C.S.R. Demokritos,
     P.O. Box 60228, GR-15310 Athens, Greece
    \label{DEMOKRITOS}}
\titlefoot{FZU, Inst. of Phys. of the C.A.S. High Energy Physics Division,
     Na Slovance 2, CZ-182 21, Praha 8, Czech Republic
    \label{FZU}}
\titlefoot{Dipartimento di Fisica, Universit\`a di Genova and INFN,
     Via Dodecaneso 33, IT-16146 Genova, Italy
    \label{GENOVA}}
\titlefoot{Institut des Sciences Nucl\'eaires, IN2P3-CNRS, Universit\'e
     de Grenoble 1, FR-38026 Grenoble Cedex, France
    \label{GRENOBLE}}
\titlefoot{Helsinki Institute of Physics and Department of Physical Sciences,
     P.O. Box 64, FIN-00014 University of Helsinki, 
     \indent~~Finland
    \label{HELSINKI}}
\titlefoot{Joint Institute for Nuclear Research, Dubna, Head Post
     Office, P.O. Box 79, RU-101 000 Moscow, Russian Federation
    \label{JINR}}
\titlefoot{Institut f\"ur Experimentelle Kernphysik,
     Universit\"at Karlsruhe, Postfach 6980, DE-76128 Karlsruhe,
     Germany
    \label{KARLSRUHE}}
\titlefoot{Institute of Nuclear Physics PAN,Ul. Radzikowskiego 152,
     PL-31142 Krakow, Poland
    \label{KRAKOW1}}
\titlefoot{Faculty of Physics and Nuclear Techniques, University of Mining
     and Metallurgy, PL-30055 Krakow, Poland
    \label{KRAKOW2}}
\titlefoot{Universit\'e de Paris-Sud, Lab. de l'Acc\'el\'erateur
     Lin\'eaire, IN2P3-CNRS, B\^{a}t. 200, FR-91405 Orsay Cedex, France
    \label{LAL}}
\titlefoot{School of Physics and Chemistry, University of Lancaster,
     Lancaster LA1 4YB, UK
    \label{LANCASTER}}
\titlefoot{LIP, IST, FCUL - Av. Elias Garcia, 14-$1^{o}$,
     PT-1000 Lisboa Codex, Portugal
    \label{LIP}}
\titlefoot{Department of Physics, University of Liverpool, P.O.
     Box 147, Liverpool L69 3BX, UK
    \label{LIVERPOOL}}
\titlefoot{Dept. of Physics and Astronomy, Kelvin Building,
     University of Glasgow, Glasgow G12 8QQ
    \label{GLASGOW}}
\titlefoot{LPNHE, IN2P3-CNRS, Univ.~Paris VI et VII, Tour 33 (RdC),
     4 place Jussieu, FR-75252 Paris Cedex 05, France
    \label{LPNHE}}
\titlefoot{Department of Physics, University of Lund,
     S\"olvegatan 14, SE-223 63 Lund, Sweden
    \label{LUND}}
\titlefoot{Universit\'e Claude Bernard de Lyon, IPNL, IN2P3-CNRS,
     FR-69622 Villeurbanne Cedex, France
    \label{LYON}}
\titlefoot{Dipartimento di Fisica, Universit\`a di Milano and INFN-MILANO,
     Via Celoria 16, IT-20133 Milan, Italy
    \label{MILANO}}
\titlefoot{Dipartimento di Fisica, Univ. di Milano-Bicocca and
     INFN-MILANO, Piazza della Scienza 3, IT-20126 Milan, Italy
    \label{MILANO2}}
\titlefoot{IPNP of MFF, Charles Univ., Areal MFF,
     V Holesovickach 2, CZ-180 00, Praha 8, Czech Republic
    \label{NC}}
\titlefoot{NIKHEF, Postbus 41882, NL-1009 DB
     Amsterdam, The Netherlands
    \label{NIKHEF}}
\titlefoot{National Technical University, Physics Department,
     Zografou Campus, GR-15773 Athens, Greece
    \label{NTU-ATHENS}}
\titlefoot{Physics Department, University of Oslo, Blindern,
     NO-0316 Oslo, Norway
    \label{OSLO}}
\titlefoot{Dpto. Fisica, Univ. Oviedo, Avda. Calvo Sotelo
     s/n, ES-33007 Oviedo, Spain
    \label{OVIEDO}}
\titlefoot{Department of Physics, University of Oxford,
     Keble Road, Oxford OX1 3RH, UK
    \label{OXFORD}}
\titlefoot{Dipartimento di Fisica, Universit\`a di Padova and
     INFN, Via Marzolo 8, IT-35131 Padua, Italy
    \label{PADOVA}}
\titlefoot{Rutherford Appleton Laboratory, Chilton, Didcot
     OX11 OQX, UK
    \label{RAL}}
\titlefoot{Dipartimento di Fisica, Universit\`a di Roma II and
     INFN, Tor Vergata, IT-00173 Rome, Italy
    \label{ROMA2}}
\titlefoot{Dipartimento di Fisica, Universit\`a di Roma III and
     INFN, Via della Vasca Navale 84, IT-00146 Rome, Italy
    \label{ROMA3}}
\titlefoot{DAPNIA/Service de Physique des Particules,
     CEA-Saclay, FR-91191 Gif-sur-Yvette Cedex, France
    \label{SACLAY}}
\titlefoot{Instituto de Fisica de Cantabria (CSIC-UC), Avda.
     los Castros s/n, ES-39006 Santander, Spain
    \label{SANTANDER}}
\titlefoot{Inst. for High Energy Physics, Serpukov
     P.O. Box 35, Protvino, (Moscow Region), Russian Federation
    \label{SERPUKHOV}}
\titlefoot{J. Stefan Institute, Jamova 39, SI-1000 Ljubljana, Slovenia
    \label{SLOVENIJA1}}
\titlefoot{Laboratory for Astroparticle Physics,
     University of Nova Gorica, Kostanjeviska 16a, SI-5000 Nova Gorica, Slovenia
    \label{SLOVENIJA2}}
\titlefoot{Department of Physics, University of Ljubljana,
     SI-1000 Ljubljana, Slovenia
    \label{SLOVENIJA3}}
\titlefoot{Fysikum, Stockholm University,
     Box 6730, SE-113 85 Stockholm, Sweden
    \label{STOCKHOLM}}
\titlefoot{Dipartimento di Fisica Sperimentale, Universit\`a di
     Torino and INFN, Via P. Giuria 1, IT-10125 Turin, Italy
    \label{TORINO}}
\titlefoot{INFN,Sezione di Torino and Dipartimento di Fisica Teorica,
     Universit\`a di Torino, Via Giuria 1,
     IT-10125 Turin, Italy
    \label{TORINOTH}}
\titlefoot{Dipartimento di Fisica, Universit\`a di Trieste and
     INFN, Via A. Valerio 2, IT-34127 Trieste, Italy
    \label{TRIESTE}}
\titlefoot{Istituto di Fisica, Universit\`a di Udine and INFN,
     IT-33100 Udine, Italy
    \label{UDINE}}
\titlefoot{Univ. Federal do Rio de Janeiro, C.P. 68528
     Cidade Univ., Ilha do Fund\~ao
     BR-21945-970 Rio de Janeiro, Brazil
    \label{UFRJ}}
\titlefoot{Department of Radiation Sciences, University of
     Uppsala, P.O. Box 535, SE-751 21 Uppsala, Sweden
    \label{UPPSALA}}
\titlefoot{IFIC, Valencia-CSIC, and D.F.A.M.N., U. de Valencia,
     Avda. Dr. Moliner 50, ES-46100 Burjassot (Valencia), Spain
    \label{VALENCIA}}
\titlefoot{Institut f\"ur Hochenergiephysik, \"Osterr. Akad.
     d. Wissensch., Nikolsdorfergasse 18, AT-1050 Vienna, Austria
    \label{VIENNA}}
\titlefoot{Inst. Nuclear Studies and University of Warsaw, Ul.
     Hoza 69, PL-00681 Warsaw, Poland
    \label{WARSZAWA}}
\titlefoot{Now at University of Warwick, Coventry CV4 7AL, UK
    \label{WARWICK}}
\titlefoot{Fachbereich Physik, University of Wuppertal, Postfach
     100 127, DE-42097 Wuppertal, Germany \\
\noindent
{$^\dagger$~deceased}
    \label{WUPPERTAL}}
\addtolength{\textheight}{-10mm}
\addtolength{\footskip}{5mm}
\clearpage
\headsep 30.0pt
\end{titlepage}
%%%%%%%%%%%%%%%%%%%%%%%%%
%
% Change for the document body
%%\pagestyle{heading} % for page numbering
\pagenumbering{arabic} % page numbering in number
\setcounter{footnote}{0} %
\large
%\linenumbers %%%CD

%\input{document.tex}    % The body of the document.

\section{Introduction}

The polarisation of tau leptons ($P_\tau$)
has been precisely measured by DELPHI \cite{ptaudelphi} and other 
LEP experiments \cite{ptaualeph},\cite{ptaul3},\cite{ptauopal} 
in Z $\rightarrow \tau^+\tau^-$ decays
during the LEP running near the Z pole (LEP1).
The measurements of the tau polarisation
allowed the LEP experiments to determine precisely the 
ratio of the electroweak axial and vector coupling constants, or 
equivalently,
the value of the effective electroweak mixing angle. Starting from 1996 
the LEP energy was increased to values significantly above
the Z resonance. In this phase, known as LEP2, the centre-of-mass 
energy $\sqrt{s}$ of the initial e$^+$e$^-$ system 
had values lying between 161 and 209 GeV.
At LEP2, due to the much reduced production cross-section,
the collected statistics of tau pairs was two orders of magnitude
smaller than at LEP1, which makes the experimental errors much larger and
therefore they have a weaker constraint on electroweak parameters.
However the determination of $P_\tau$ at the world's highest 
energies of e$^+$e$^-$ annihilation is still important
for the search for deviations from the Standard Model predictions
(e.g. existence of a Z$^\prime$ boson).

In this Letter we present the determination of the polarisation of tau
leptons produced in e$^+$e$^-$ annihilations at energies between
183 and 209 GeV. The data were collected in the DELPHI experiment during
1997-2000. 
The data collected during 1996 were not included 
because of the low integrated luminosity recorded.
The analysis was based on the sample of tau pairs selected for
the measurement of the production cross-section and forward-backward
asymmetry \cite{lep2}. 

At LEP the tau leptons produced in pairs have opposite helicity.
Throughout this paper we refer to the helicity and polarisation
of $\tau^-$. The average tau polarisation $P_\tau$ is defined 
as the relative excess of the right-handed $\tau^-$ over 
the left-handed ones: 
\begin{equation}
P_\tau = \frac{N_R - N_L}{N_R + N_L} .
\end{equation}

\noindent
The polarisation dependence on the tau production
angle was not measured because of too low statistics
of the backward tau production at LEP2. In this Letter
$P_\tau$ denotes the average polarisation over all tau 
production angles.

At LEP2 a significant fraction 
of fermion pairs was produced in the {\it radiative return} process,
when the annihilation energy was reduced to the Z resonance
region by the radiation of a hard photon
from the initial state.
To ensure that the e$^+$e$^-$ annihilation occurred
at high energy 
the reconstructed centre-of-mass energy of the tau pair
($\sqrt{s^\prime}$) was required to be close to the nominal LEP energy:
$\sqrt{s^\prime/s} > 0.92$. The determination of $\sqrt{s^\prime}$ was based
on the measured directions of the jets of tau decay products. The
procedure of the tau pair selection and $\sqrt{s^\prime}$ determination is
described in detail in \cite{lep2}. The detector calibration 
and systematic error determination was also largely based on
the procedures described in \cite{lep2}.
A detailed description of the DELPHI detector and its performance
can be found in \cite{delphidet} and
\cite{delphiperf}.

The signal process e$^+$e$^- \rightarrow \tau^+\tau^-$ was simulated 
using the KK Monte Carlo generator \cite{KK}, 
while tau decays were handled by TAUOLA 2.6 \cite{tauola}.
The main background processes
were simulated using the following generators: BHWIDE \cite{bhwide}
for e$^+$e$^- \rightarrow$ e$^+$e$^-$; 
KK for e$^+$e$^- \rightarrow \mu^+\mu^-$; KK and PYTHIA \cite{pythia}
for  e$^+$e$^- \rightarrow q\overline{q}$; WPHACT \cite{wphact}
for e$^+$e$^- \rightarrow $ W$^+$W$^-$, e$^+$e$^- \rightarrow $ ZZ and
e$^+$e$^- \rightarrow $ Ze$^+$e$^-$; 
BDK/BDKRC \cite{bdk} for $\gamma\gamma \rightarrow $ e$^+$e$^-$,
$\gamma\gamma \rightarrow \mu^+\mu^-$ and
$\gamma\gamma \rightarrow \tau^+\tau^-$; and PYTHIA for 
$\gamma\gamma \rightarrow q\overline{q}$. The generated events were
passed through the full chain of the detector simulation, event
reconstruction and data analysis.
The procedure of the Monte Carlo simulation 
of the DELPHI detector is described in \cite{delphiperf}.

\section{Event selection}

The determination of the average tau polarisation was based
on the inclusive selection of one-prong hadronic decays
of tau leptons. Leptonic and multi-track tau decays
were not used because of their very low sensitivity 
to the polarisation.
The method closely followed the one
developed for the LEP1 analysis \cite{ptaudelphi}, with modifications
necessary to take into account the increased centre-of-mass energy
and the lower number of tau pairs observed at LEP2.
The charged particles in each preselected event
were combined into two jets using the PYCLUS algorithm
\cite{pythia}. The most energetic charged
particle ({\it leading track}) was determined for each jet and all
tracks and electromagnetic showers within a 30$^\circ$ cone around each
leading track were assumed to originate from the decay of the tau lepton. 
The two tau decay candidates in each event were then analysed separately. 
An important quantity for this analysis, the {\it visible invariant mass}
($M_{VIS}$), was calculated for each tau decay candidate using all 
charged particles
(assumed to be pions) and all photons, i.e. electromagnetic showers
with energy above 0.5 GeV unassociated with a charged particle.

The one-prong hadronic tau decays were selected using the following
procedure. The leading track had to be reconstructed within the 
barrel part of the DELPHI detector (polar angle\footnote{The DELPHI 
coordinate system is a right-handed 
system with the $z$-axis collinear with the incoming electron beam,
the $x$-axis pointing to the centre of the LEP accelerator and the 
$y$-axis vertical. The polar angle $\theta$ is with reference to the
$z$-axis, and $\phi$ is the azimuthal angle in the $x,y$ plane.} 
range $41^\circ < \theta < 139^\circ$). 
Tracks close to the DELPHI middle plane
($88.5^\circ < \theta < 91.5^\circ$) were excluded. Tau decay candidates
in which the leading track extrapolation passed closer than
0.3$^\circ$ from the centre of 
a $\phi$-crack
of the barrel electromagnetic calorimeter (HPC) were also excluded.
The leading track had to be the only track originating from the tau
decay, with the exception of the tracks that were reconstructed 
as an e$^+$e$^-$ pair from a conversion (such pairs were treated as photons
in the analysis). 
The procedure of the conversion 
reconstruction is described in \cite{delphiperf}.

Tau decays to electrons of relatively low energy were rejected
by the requirement that the measured dE/dx losses of the 
charged particle
as measured in the Time Projection Chamber (TPC)
did not exceed the value expected for a pion by more than 2 standard
deviations. Electrons of higher energies were suppressed by 
requiring that at least one of the two following conditions was satisfied:
either the energy deposition in the HPC associated
to the charged particle
had to be less than 10 GeV or the associated deposition
beyond the first layer of the Hadron Calorimeter (HCAL) had to be
greater than 0.5 GeV. In the cases where a dE/dx measurement
was not available, the event was rejected if the particle momentum
was in the range below 10 GeV/c for which the HPC 
energy measurement is less precise. 

The tau decays involving muons were suppressed 
by the requirement that no hits in the muon chambers were associated
to the charged particle
by the standard DELPHI procedure of muon
identification \cite{delphiperf}. For the tau decay candidates
with low visible invariant mass ($M_{VIS} < 0.3$ GeV/c$^2$) an additional 
muon-suppression was applied: the average measured energy
deposition per HCAL layer
associated to the charged particle
had to be inconsistent with a
minimum ionizing particle, namely it 
had to lie outside the range 0.5 to 1.5 GeV.

During the whole period of data taking in 2000 
the performance of one of the 12 sectors
of the DELPHI TPC was unstable.
The good performance of the TPC is crucial for this analysis,
in particular for the dE/dx measurements. Therefore for the
data taken in 2000 the selection procedure was modified. 
A tau decay candidate was rejected if the leading track 
was reconstructed within the faulty TPC sector or close
to it (within 10$^\circ$ in azimuthal angle). This reduced 
the selection efficiency for the 2000 data by \mbox{approximately 10\%.}

\begin{figure}
\begin{center} \mbox{\epsfig{file=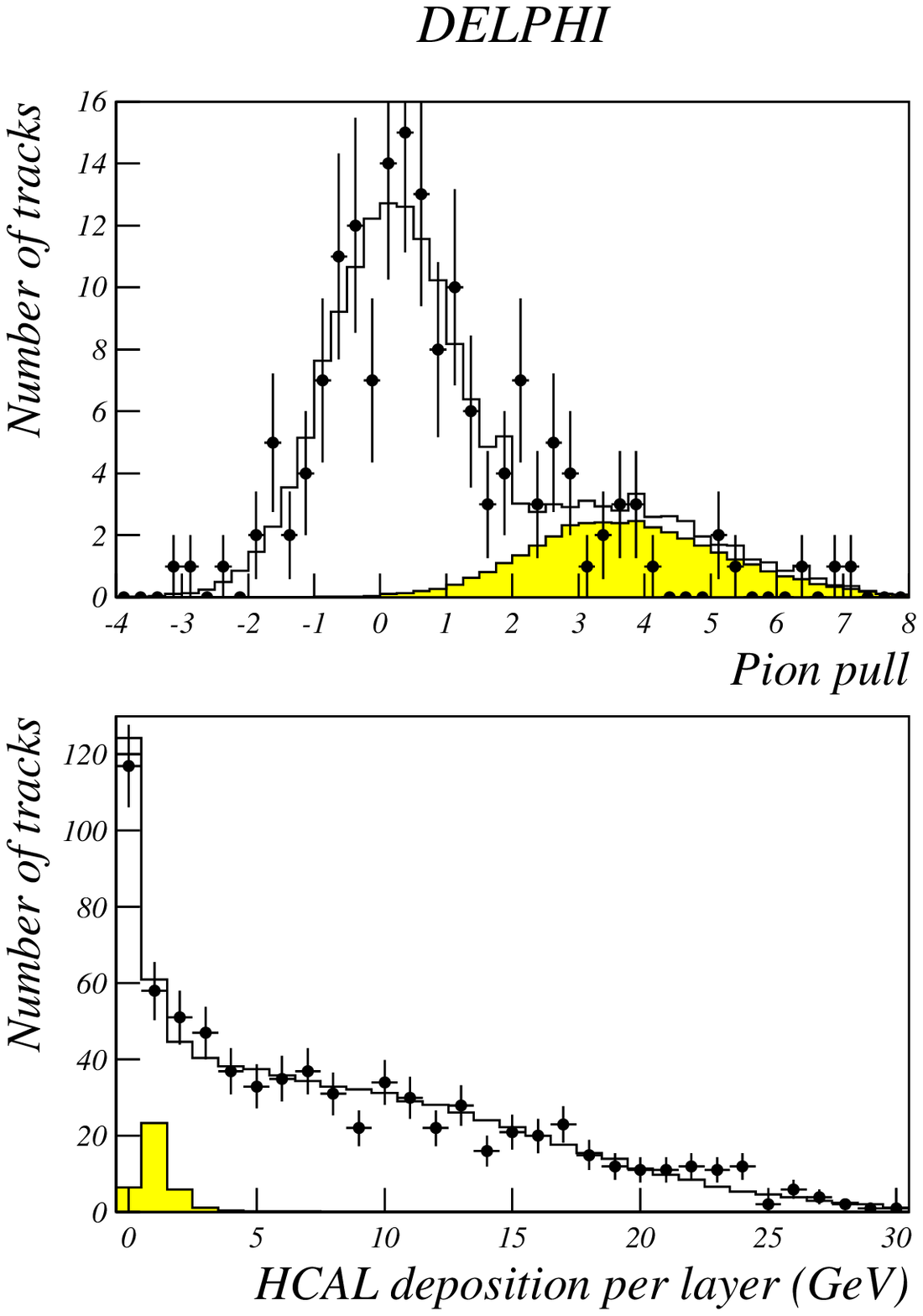,width=14cm}}
\end{center}
\caption{Top: distribution of the dE/dx pion hypothesis pull.
The grey/yellow area shows the contribution expected from electrons. Bottom: 
distribution of the average energy deposition per HCAL layer. The 
grey/yellow area
shows the contribution from muons. In both plots the real data
are represented by points and the solid lines show the simulation.
}
\label{fig:var}
\end{figure}

Two of the event selection variables are illustrated in 
Fig. \ref{fig:var}.
The upper plot shows the distribution of the so-called
``dE/dx pull'' for the pion hypothesis, i.e. the difference between 
the measured dE/dx losses of the charged particle and the value expected
for a pion, expressed in number of standard deviations 
(see \cite{ptaudelphi} for the exact definition),
for particles with momentum below 12 GeV/c.
The lower plot shows the distribution of the average energy deposition
per HCAL layer associated to the charged particle. The grey areas 
in Fig. \ref{fig:var}
show the background most relevant to the variable shown.
The data shown in Fig. \ref{fig:var} represent
the full statistics of 1997-2000.

In total, 624 hadronic tau decay candidates were selected 
from the 1997-2000 data.
The details of the selection for each year of data taking 
are summarised in Table \ref{tab:del}. The efficiency values are given
within the polar angle acceptance. The efficiency drop in 2000
is due to the rejection of 
particles crossing the faulty TPC sector.
The non-tau background consisted mainly of e$^+$e$^- \rightarrow$ e$^+$e$^-$,
e$^+$e$^- \rightarrow $ W$^+$W$^-$ and e$^+$e$^- \rightarrow $ 
Ze$^+$e$^-$ events (in approximately equal fractions).
The selection efficiency and the background level were
determined from the simulation. Small corrections (typically 10\%)
were applied to the residual non-tau background
to account for the differences between data and simulation.
The procedure for this correction is described in \cite{lep2}.

\begin{table}[h]
\begin{center}
\begin{tabular}{|c|c|c|c|c|}
\hline
Year & 1997 & 1998 & 1999 & 2000 
\\ \hline\hline
Mean $\sqrt{s}$ (GeV) & 183 & 189 & 198 & 206 
\\ \hline
Integrated luminosity (pb$^{-1}$)& 52 & 153 & 224 & 217
\\ \hline\hline
Number of selected & 82 & 231 & 305 & 254
\\
tau pairs (in barrel) & & & &
\\ \hline
Number of selected & 56 & 159 & 234 & 175
\\
hadronic tau decays  & & & & 
\\ \hline
Hadronic selection efficiency (\%) & 77.3 & 77.1& 77.1 & 70.3
\\ \hline 
Non-tau background (\%)& 4.6 & 3.8 & 4.7 & 4.4
\\ \hline 
Tau leptonic decay background (\%) & 3.3 & 3.4 & 3.2 & 3.3
\\ \hline 
Fraction (\%) of events & & & & 
\\
with $\sqrt{s^\prime/s} < 0.92$ & 5.3 & 4.9 & 4.9 & 5.0
\\ \hline 
\end{tabular}
\caption{Results of the tau hadronic decay selection. 
}
\label{tab:del}
\end{center}
\end{table}

\begin{figure}
\begin{center} \mbox{\epsfig{file=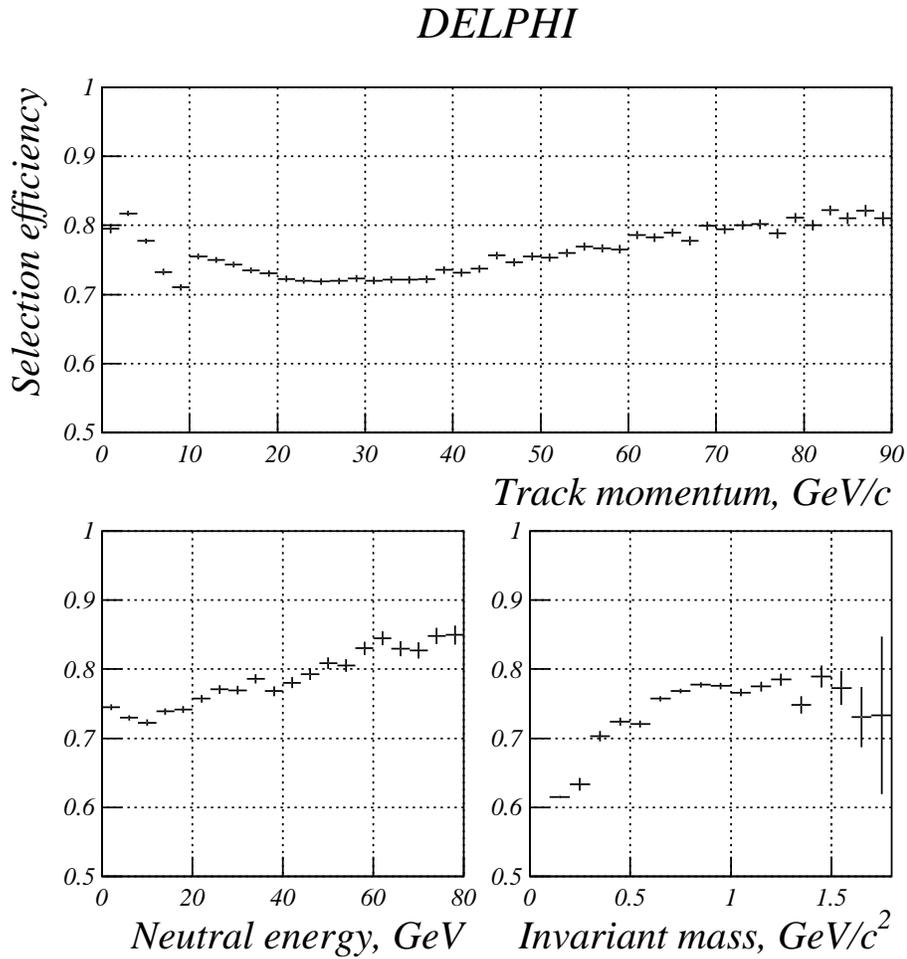,width=14cm}}
\end{center}
\caption{Efficiency of the 1-prong hadronic tau decay selection
versus the kinematic variables: momentum
of the charged particle; total energy of photons; and the visible
invariant mass of tau decay products.
The error bars represent the statistical 
uncertainty of the simulation sample. 
The step at 10 GeV/c (upper plot) is caused by the 
rejection of tracks without dE/dx measurement.
}
\label{fig:eff}
\end{figure}

Fig. \ref{fig:eff} shows the dependence of the selection efficiency
on the variables which are sensitive to the tau polarisation:
momentum of the charged particle; 
total energy of photons from the tau decay; 
and $M_{VIS}$. The step at 10 GeV/c momentum
is caused by the different treatment of the tracks without dE/dx 
measurement. The drop of efficiency at low invariant masses
is due to the tighter muon rejection in this region.
In general, the efficiency is relatively flat, which is important
for an unbiased polarisation measurement.

\begin{figure}
\begin{center} \mbox{\epsfig{file=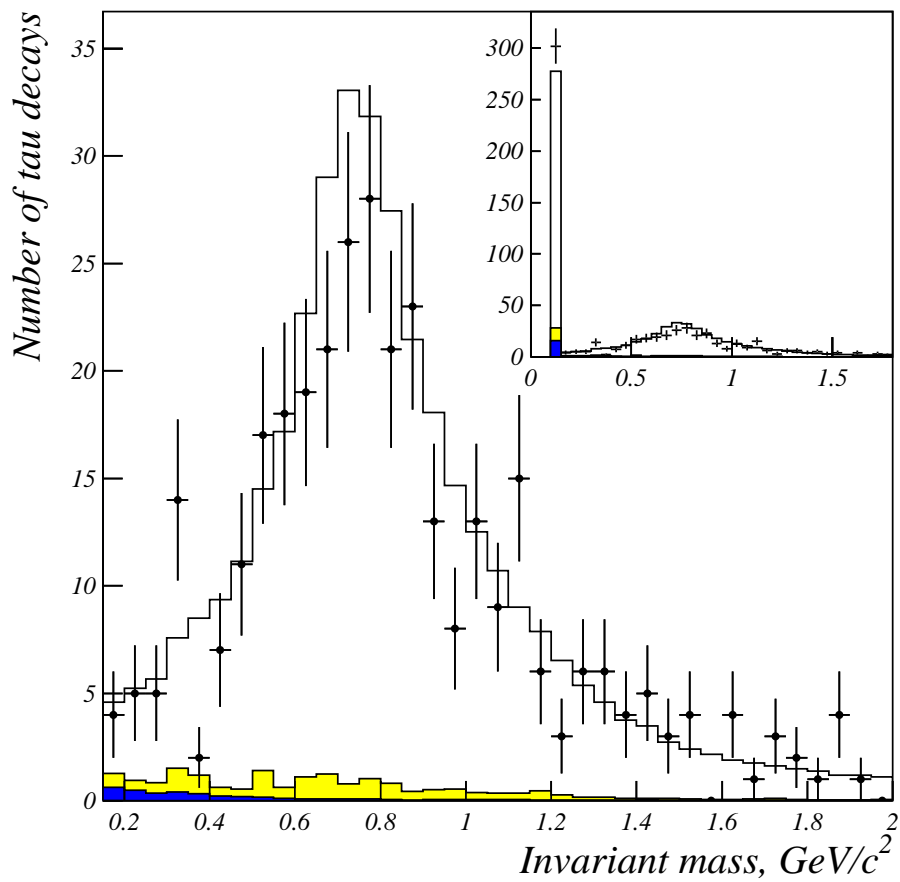,width=14cm}}
\end{center}
\caption{Distribution of the visible invariant mass.
The points represent data, the solid line is the simulation,
and the grey/yellow and black/blue areas show the contributions 
respectively from non-tau 
background and from leptonic tau decays.
The main plot and the inset show the same distributions
in different scale.
}
\label{fig:minv}
\end{figure}

The distribution of the visible invariant mass for the selected
decays is shown in Fig. \ref{fig:minv}. The main plot does not show 
the first bin corresponding to $\tau \rightarrow \pi\nu$ decays.
The same distribution, including the first bin, is shown in the inset.

\section{Determination of the tau polarisation}

The selected sample mainly consisted of the decays 
$\tau \rightarrow \pi\nu$, $\tau \rightarrow \rho\nu$ and
$\tau \rightarrow a_1\nu$. Mixing the different decay modes in the
inclusive sample reduces the analysis sensitivity to the polarisation.
In order to improve the sensitivity
the analysis was performed in three bins of the visible
invariant mass: \mbox{$0 < M_{VIS} < 0.3$ GeV/c$^2$}, dominated by 
$\tau \rightarrow \pi\nu$ (59\%); 
\mbox{0.3 GeV/c$^2 < M_{VIS} < 0.8$ GeV/c$^2$},
dominated by $\tau \rightarrow \rho\nu$ (78\%); and 
\mbox{0.8 GeV/c$^2 < M_{VIS} < 2.0$ GeV/c$^2$}, populated by 
$\tau \rightarrow \rho\nu$ (61\%) and $\tau \rightarrow a_1\nu$ (34\%). 
The total numbers of decays selected
in each bin of $M_{VIS}$ were 316, 153 and 155 respectively.

As in the LEP1 analysis \cite{ptaudelphi} the extraction of the 
tau polarisation
was based on reconstruction of the two kinematic variables
characterizing the tau decay: $\Theta$, the angle in the
$\tau$ rest frame between the
momenta of $\tau$ and $h$ for $\tau \rightarrow h\nu$ decays; 
and $\Psi$ which, in the case of $\tau \rightarrow \rho\nu$ decay,
is the angle of the emission of the pions in the $\rho$ rest frame.
The angle $\Theta$ was reconstructed as

\begin{equation}
\cos{\Theta} = \frac{2p_h/p_\tau-1-m_h^2/m_\tau^2}{1-m_h^2/m_\tau^2},
\end{equation}

\noindent where $p_h$ is the momentum of the hadronic system
produced in the tau decay (vector sum of the momenta of the
reconstructed tau decay products) and 
$m_h$ is the mass of the hadronic system (experimentally
reconstructed as $M_{VIS}$).
The tau lepton momentum $p_\tau$ was estimated 
from the directions of the jets of the tau decay products
using the same method as for the determination of the $\sqrt{s^\prime}$ 
value (see \cite{lep2} for a detailed explanation). 
The uncertainty of the $p_\tau$ determination was 
approximately 1.5\%, mainly due to the unknown energies and directions
of the neutrinos produced in the tau decays.
The angle $\Psi$ was determined from 

\begin{equation}
\cos{\Psi} = \frac{E_{ch}-E_{neu}}{E_{ch}+E_{neu}},
\end{equation}

\noindent where $E_{ch}$ and $E_{neu}$ are the energy of the charged
particle and the total energy of the photons from the tau decay.
For visible invariant masses above 0.3 GeV/c$^2$ the range
$\cos{\Theta}>0.8$ was rejected because it was dominated 
by events with wrongly reconstructed kinematics.

The value of the tau polarisation was extracted 
from a binned likehood fit to the observed distributions
of $\cos{\Theta}$ and $\cos{\Psi}$
by the simulation expectation $f_{MC}$
with the $P_\tau$ value being a free fit parameter:

\begin{equation}
f_{MC} = f_{bg} + 
R\cdot\left (\frac{1-P_\tau}{1-P_0}f_L + \frac{1+P_\tau}{1+P_0}f_R\right ),
\end{equation}

\noindent
where $f_{bg}$, $f_L$ and $f_R$ are the contributions from external (non-tau)
background and from decays of left- and right-handed 
tau leptons,
and $P_0$ is the generator level tau polarisation in the 
simulated tau pair sample.
The external background contribution was normalized to the luminosity.
The factor R normalizes the number of events in the simulated tau signal 
to the real data after external background subtraction:

\begin{equation}
R\cdot N^{\tau}_{MC} = N_{data} - N_{bg},
\end{equation}

\noindent
where $N_{data}$ is the number of observed events,
$N^{\tau}_{MC}$ is the number of simulated signal
events, and $N_{bg}$ is the non-tau background 
predicted by simulation.
Such a fit automatically takes into account the bias due to different
selection efficiencies for different tau helicities.
It does not depend on the tau polarisation 
in the simulated tau pair sample.

The tau polarisation was extracted separately for each year of the data taking.
The two-dimensional distributions of $\cos{\Theta}$ versus 
$\cos{\Psi}$ were fitted
simultaneously in the three bins of the invariant mass.
For the first bin of invariant mass only the one-dimensional distribution of 
$\cos{\Theta}$ was used
because this bin is dominated by decays to pions where $\Psi$ 
has no meaning. 
The results of the fits are presented in Table \ref{years},
together with their average. The Table also shows the statistical
uncertainty of the $P_\tau$ determination and the uncertainties 
associated with the finite statistics of the simulated events.
This Table shows the results obtained from the fit before applying the
corrections discussed in the next section.
Despite the apparent energy dependence, the results are consistent
with being constant with energy. The $\chi^2/n.d.f.$ 
for a constant value is 5.0/3.

\begin{table}[h]
\begin{center}
\begin{tabular}{|c|c|c|c|}
\hline
Year & $\tau$ polarisation & stat. error & Simulation stat. error
\\ \hline\hline
1997 & -0.61 & 0.34 & 0.015
\\ \hline
1998 & -0.41 & 0.21 & 0.009
\\ \hline
1999 & -0.01 & 0.20 & 0.009
\\ \hline 
2000 & +0.11 & 0.24 & 0.010
\\ \hline  \hline 
Average & -0.176 & 0.117 & 0.005
\\ \hline 
\end{tabular}
\caption{Values of the tau polarisation determined from each year's data,
and their average. Also shown are the statistical errors from the fits
and the uncertainty due to the limited statistics of the simulation
samples.
}
\label{years}
\end{center}
\end{table}

As a cross-check, the result was also obtained with a single fit
to the whole data sample (1997-2000).
The Monte Carlo samples were combined with weights proportional 
to the integrated luminosity of the respective year.
The result of this fit was -0.140$\pm$0.123,
which is less than one standard deviation from the average in
Table \ref{years} (allowing for the high statistical correlation
between both values).
The average of the year-by-year measurements was chosen to produce the final
result because the year-specific Monte Carlo samples should
better reproduce differences in detector performance and 
calibration in the different periods of data taking.

\begin{figure}
\begin{center} \mbox{\epsfig{file=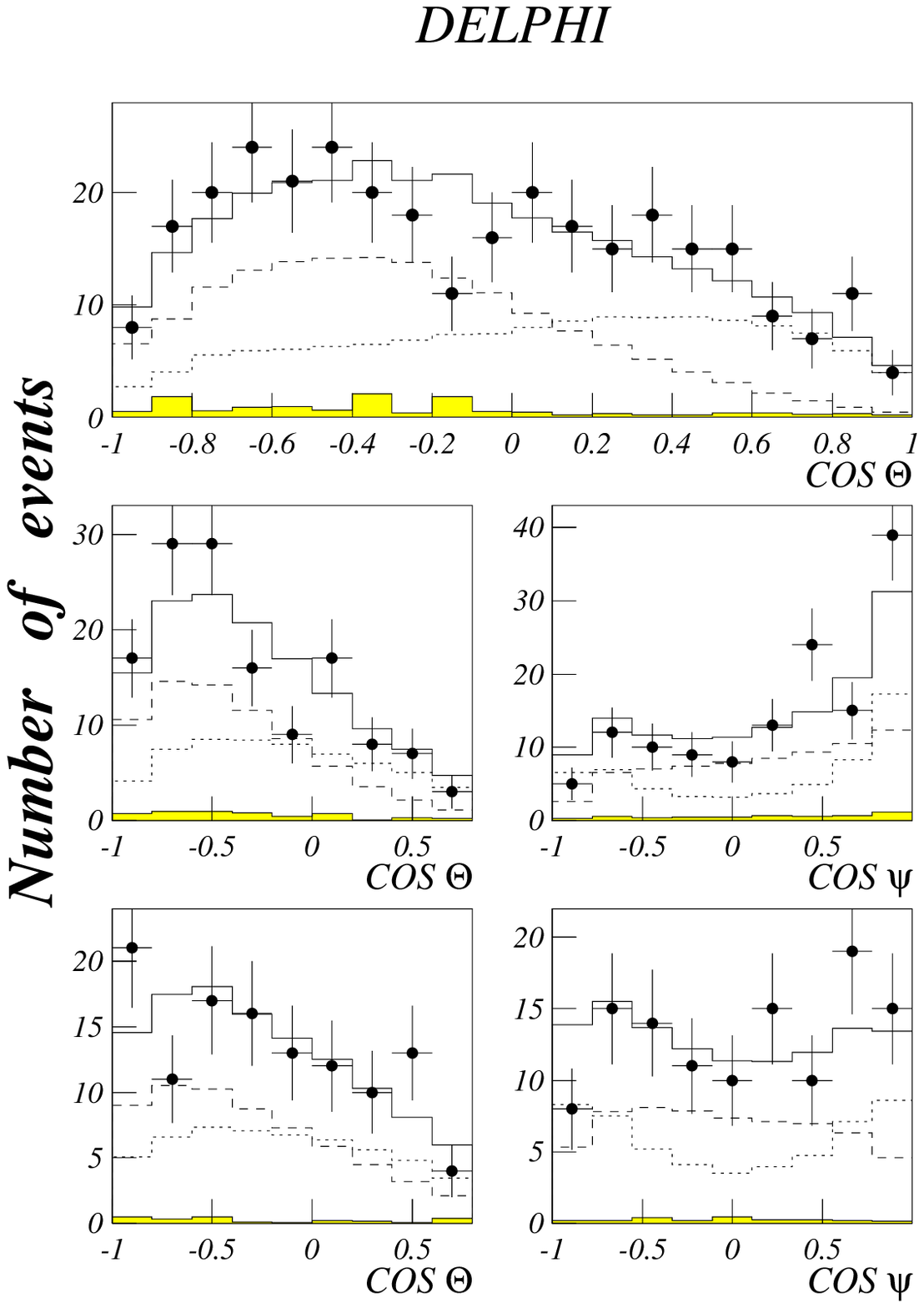,width=14cm}}
\end{center}
\caption{The results of the tau polarisation fit for different 
bins of $M_{VIS}$: 0 -- 0.3 GeV/c$^2$ (upper plot), 
0.3 -- 0.8 GeV/c$^2$ (middle plots)
and 0.8 -- 2.0 GeV/c$^2$ (lower plots). 
The points represent data, the grey/yellow areas show the non-tau background,
the dashed and dotted lines show the contributions
from the decays of left- and right-handed tau leptons,
and the solid lines show the total prediction of simulation.
}
\label{fig:fit1}
\end{figure}

The results of the fit are illustrated in Fig. \ref{fig:fit1}
which shows the distribution of $\cos{\Theta}$ for the first bin
of invariant mass and 
one-dimensional projections of the fitted two-dimensional
distributions for other invariant masses. 
Combined data of all years are shown by the points with error
bars and the
simulation is shown by the solid lines.
The distributions for simulated tau decays are shown with the polarisation
value which was obtained in this study. The contributions
from the decays of left- and right-handed tau leptons
are shown by the dashed and dotted lines respectively. The contribution
of the non-tau background is shown as a grey/yellow area.

\section{Corrections and systematic errors}

A small correction had to be applied to the measured polarisation
to subtract the contribution of the {\it feed-through} events,
i.e. the events which have true values of $\sqrt{s^\prime/s}$ below 0.92
although they pass the experimental cut of $\sqrt{s^\prime/s} > 0.92$
(see Table \ref{tab:del}).
After such a correction the measured polarisation represents
the average polarisation of tau leptons produced at the actual
annihilation energies above $0.92\cdot\sqrt{s}$. 
The value of the correction
depends on the measured polarisation. Since the results
from individual years (Table \ref{years}) are consistent 
with each other, and the polarisation dependence on energy
is weak,
we apply to the results of all years 
the same global correction calculated using the KK generator
for the average measured polarisation.
The value of the correction was found to be +0.004.

This method of tau polarisation measurement 
depends on a good description of the data 
by the simulation. Therefore an extensive study 
of the simulation quality has been performed
using high purity test samples selected from data and
simulation. The uncertainties of such checks
(dominated by the statistics of test samples 
selected from data) were converted into the 
systematic uncertainty of the polarisation measurement.
To reduce the effect of statistical fluctuations,
the test samples were selected from the combined
1997-2000 data. The systematic uncertainties therefore
were common to all years of the data taking.
Most of the corrections and corresponding systematic uncertainties
were propagated from the study of tau pair
production, see \cite{lep2}. Some of these 
correspond to small corrections applied to variables
at the very beginning of the analysis, before the tau pair selection,
such as the correction to the measured dE/dx (see below),
which are therefore already included in the results
of Table \ref{years}. In other  cases they had to be calculated
as corrections to the results and have to be added to those.
In these cases the correction values are given below.
A conservative approach was followed, applying a correction and 
uncertainty even in the cases where the correction was consistent with zero.

The dE/dx measurements were calibrated using test samples of 
muons from the processes $\gamma\gamma \rightarrow \mu^+\mu^-$,
e$^+$e$^- \rightarrow \mu^+\mu^-$ and Z $\rightarrow \mu^+\mu^-$
(the latter were produced during the short periods of LEP 
running near the Z pole in 1997-2000). Both the dE/dx mean value and 
the measurement resolution were calibrated and a small momentum-dependent
correction was applied. The uncertainty due to the 
calibration gave rise to an uncertainty of $\pm 0.017$ in $P_\tau$.

The measurement of photon energy was important for the reconstruction
of the tau hadronic decay kinematics. The electromagnetic 
energy scale was checked using a sample of electrons 
from $\gamma\gamma \rightarrow $ e$^+$e$^-$, 
e$^+$e$^- \rightarrow$ e$^+$e$^-$ and Z $\rightarrow $ e$^+$e$^-$
events. A correction of $-0.010 \pm 0.010$ to the tau polarisation
was found to be necessary.

The redundancy between the HPC and HCAL was used to estimate from the data
the efficiency of the ``HPC or HCAL'' cut which rejects electrons.
The momentum dependence of the cut efficiency was found to be slightly
different in data and in simulation. A correction of $+0.018 \pm 0.022$
was applied to the $P_\tau$ value. 

From the data/simulation comparison for the distribution of the number 
of reconstructed photons in tau hadronic decays it was found that the 
photon reconstruction efficiency was well described by the simulation.
The uncertainty of this check resulted in a $\pm 0.016$ uncertainty
on the $P_\tau$ value.

The efficiency of the muon rejection cuts was checked using 
the redundancy of the HCAL and the muon chambers. 
The muon chamber efficiency was slightly (4-7\%) higher in simulation
than in the data. The discrepancy was corrected by randomly removing
a fraction of muon chamber hits in simulation. 
An uncertainty of $\pm 0.012$ on $P_\tau$ was associated
with this correction.

The systematic uncertainty associated with the residual background 
level was determined by varying the background by $\pm 20\%$.
The size of this variation was estimated from the small residual
data/simulation disagreements in the shapes of background-sensitive 
distributions. The statistical contribution
from the number of simulated background events 
was negligible.
The resultant $P_\tau$ uncertainty was $\pm 0.014$ for 
the background from tau leptonic decays and $\pm 0.004$ for the
non-tau background.

Other possible systematic errors were estimated from variations
of the selection cuts and from changing the choice of binning
of the variables used in the fit of the tau polarisation.

The full list of systematic errors is summarized in Table \ref{tab:sys}.
Where necessary the corrections to the measured tau polarisation
are also given.

\begin{table}[h]
\begin{center}
\begin{tabular}{|c|c|c|}
\hline
Source& $P_\tau$ uncertainty & $P_\tau$ correction
\\ \hline\hline
dE/dx calibration & 17 & --
\\ \hline
$E_\gamma$ scale & 10 & -10
\\ \hline
``HPC or HCAL'' efficiency & 22  & +18
\\ \hline
$\gamma$ reconstr. efficiency & 16 & --
\\ \hline
Muon chamber efficiency & 12 & --
\\ \hline
Internal background & 14 & --
\\ \hline
External background  & 4 & --
\\ \hline
Variation of cuts & 9 & --
\\ \hline
Binning choice& 20 & --
\\ \hline
Simulation statistics & 5 & --
\\ \hline
Feed-through & -- & +4
\\ \hline \hline
Total & 45 & +12
\\ \hline
\end{tabular}
\caption{Summary of systematic uncertainties and
corrections to the tau polarisation. All values are in units of $10^{-3}$.
}
\label{tab:sys}
\end{center}
\end{table}

\section{Results and conclusions}

\begin{figure}
\begin{center} \mbox{\epsfig{file=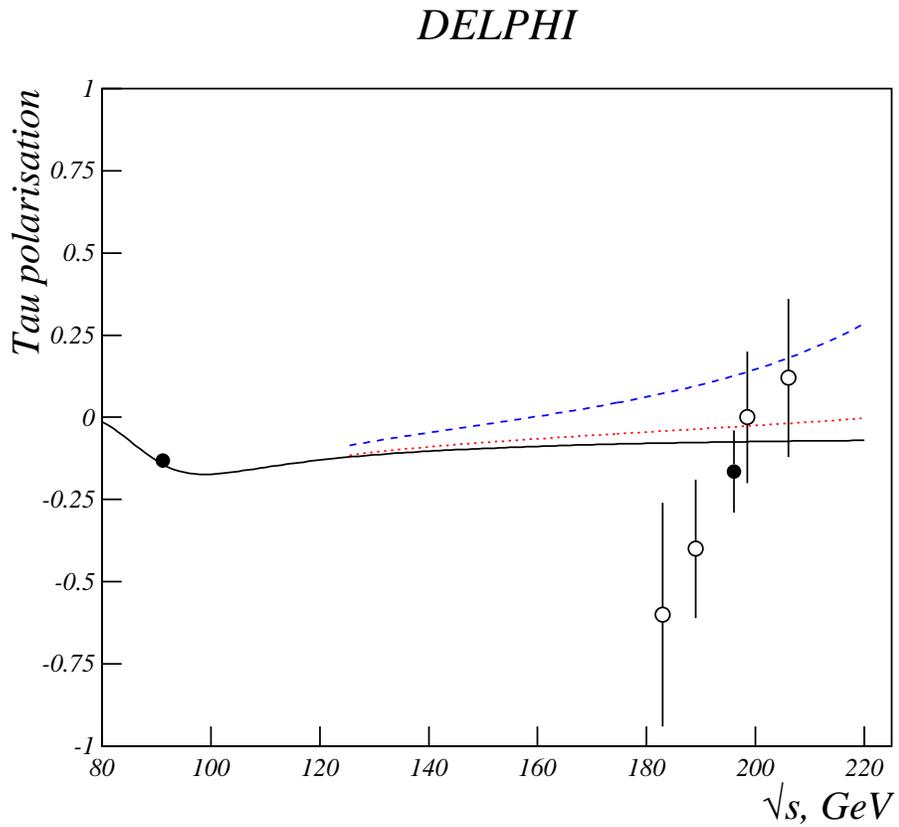,width=14cm}}
\end{center}
\caption{
Energy dependence of the tau polarisation. Black circles
show the average DELPHI measurements at LEP1 and LEP2. The white
circles are the DELPHI measurements at different LEP2 energies.
The solid line shows the Standard Model prediction 
(ZFITTER 6.36).
The dashed  and dotted lines show 
the effects from 300 and 455 GeV/c$^2$ Z$^\prime$ bosons respectively.
}
\label{fig:lep}
\end{figure}

As can be seen in Table \ref{tab:sys} the total correction 
that has to be applied to the observed value of the tau polarisation
is +0.012. After taking into account this correction the average tau 
lepton polarisation measured at LEP2 is

$$
P_\tau = -0.164 \pm 0.117 \pm 0.045,
$$

\noindent
where the first uncertainty is statistical and the second is systematic. 
Fig. \ref{fig:lep} presents the centre-of-mass energy dependence
of the tau polarisation measured by the DELPHI experiment. The plot
shows the LEP1 precision measurement and the measurements 
at the four LEP2 energies. Also shown is the average LEP2 value
which corresponds to a luminosity-weighted mean collision energy of
197 GeV. The solid curve shows the 
theoretical predictions calculated using the ZFITTER version 6.36
package \cite{zfitter}.
The calculations used the Standard Model
parameters determined at LEP1 and SLD \cite{lep1sld}.
Two other curves illustrate the effect of the existence
of a Z$^\prime$ boson in left-right models, assuming
$\alpha_{LR} = \sqrt{2/3}$ \cite{zprime}. The dashed curve
corresponds to $M_{Z^\prime} = $ 300 GeV/c$^2$ and the dotted
curve represents the DELPHI limit $M_{Z^\prime} = $ 455 GeV/c$^2$
derived from the measured fermion pair production cross-section
and charge asymmetry \cite{lep2}.

In summary, we have measured the polarisation of tau leptons 
produced at the world's 
highest e$^+$e$^-$ annihilation energy. The values measured
at different energies between 183 and 209 GeV are consistent. 
The average tau polarisation value $-0.164 \pm 0.125$ 
is consistent with the Standard Model prediction of -0.075
at the corresponding mean energy of 197 GeV.
This measurement excludes positive values of the tau polarisation
at the 90\% confidence level.

%\input{acknow.tex}
%         Modified on 04-06-1999 by dimartino
%-------------------------------------------------------------------
\subsection*{Acknowledgements}
\vskip 3 mm
We are greatly indebted to our technical 
collaborators, to the members of the CERN-SL Division for the excellent 
performance of the LEP collider, and to the funding agencies for their
support in building and operating the DELPHI detector.\\
We acknowledge in particular the support of \\
Austrian Federal Ministry of Education, Science and Culture,
GZ 616.364/2-III/2a/98, \\
FNRS--FWO, Flanders Institute to encourage scientific and technological 
research in the industry (IWT) and Belgian Federal Office for Scientific, 
Technical and Cultural affairs (OSTC), Belgium, \\
FINEP, CNPq, CAPES, FUJB and FAPERJ, Brazil, \\
%Czech Ministry of Industry and Trade, GA CR 202/99/1362,\\
%Ministry of Education of the Czech Republic LA134,\\
Ministry of Education of the Czech Republic, project LC527, \\
Academy of Sciences of the Czech Republic, project AV0Z10100502, \\
Commission of the European Communities (DG XII), \\
Direction des Sciences de la Mati$\grave{\mbox{\rm e}}$re, CEA, France, \\
Bundesministerium f$\ddot{\mbox{\rm u}}$r Bildung, Wissenschaft, Forschung 
und Technologie, Germany,\\
General Secretariat for Research and Technology, Greece, \\
National Science Foundation (NWO) and Foundation for Research on Matter (FOM),
The Netherlands, \\
Norwegian Research Council,  \\
State Committee for Scientific Research, Poland, SPUB-M/CERN/PO3/DZ296/2000,
SPUB-M/CERN/PO3/DZ297/2000, 2P03B 104 19 and 2P03B 69 23(2002-2004)\\
FCT - Funda\c{c}\~ao para a Ci\^encia e Tecnologia, Portugal, \\
Vedecka grantova agentura MS SR, Slovakia, Nr. 95/5195/134, \\
Ministry of Science and Technology of the Republic of Slovenia, \\
CICYT, Spain, AEN99-0950 and AEN99-0761,  \\
The Swedish Research Council,      \\
Particle Physics and Astronomy Research Council, UK, \\
Department of Energy, USA, DE-FG02-01ER41155, \\
EEC RTN contract HPRN-CT-00292-2002. \\

%=========================================================================%

\end{document}